\documentclass[aps,pra,showpacs,twoside,twocolumn,10pt]{revtex4-2}
\usepackage[colorlinks=true, citecolor=blue, urlcolor=blue ]{hyperref}
\usepackage{epsfig,newlfont,amssymb,amsfonts,amsmath,bm,subfigure,palatino,mathtools,amsthm,braket,soul,enumitem,graphics,graphicx,times,physics, multirow, makecell}

\usepackage[normalem]{ulem}
\usepackage{lmodern}
\usepackage[table,xcdraw]{xcolor}

\begin{document}

\title{Detecting exceptional points through dynamics in non-Hermitian systems }
\author{Keshav Das Agarwal, Tanoy Kanti Konar, Leela Ganesh Chandra Lakkaraju, Aditi Sen(De)}

\affiliation{Harish-Chandra Research Institute, A CI of Homi Bhabha National Institute,  Chhatnag Road, Jhunsi, Allahabad - 211019, India}

\begin{abstract}

Non-Hermitian rotation-time reversal ($\mathcal{RT}$)-symmetric spin models possess two distinct phases, the unbroken phase in which the entire spectrum is real and the broken phase which contains complex eigenspectra, thereby indicating a transition point, referred to as an exceptional point. We report that the dynamical quantities, namely the short- and long-time average of the Loschmidt echo which is the overlap between the initial and the final states, and the corresponding rate function, can faithfully predict the exceptional point. In particular, when the initial state is prepared as the ground state in the unbroken phase of the non-Hermitian Hamiltonian and the system is quenched to either the broken or unbroken phase, we analytically demonstrate that the rate function and the average Loschmidt echo can distinguish between the quench that occurred in the broken or the unbroken phase for the nearest-neighbor non-Hermitian $XY$ model with uniform and alternating magnetic fields, thereby indicating the exceptional point. Furthermore, we exhibit that such quantities are capable of identifying the exceptional point even in models like the non-Hermitian short- and long-range $XYZ$ model with magnetic field which can only be solved numerically, thereby establishing it as detection criteria for recognizing exceptional points.
 
\end{abstract}
\maketitle

\section{Introduction}

Non-Hermitian systems have recently attracted significant interest as a result of the intricate extension of quantum mechanics and the advancements made in quantum technologies, such as quantum sensing \cite{jan_pra_2016,chen_nature_2017,hodaei_nature_2017}, quantum metrology \cite{chen_njp_2019}, and state tomography \cite{naghiloo_nature_2019}.
Importantly, several characteristics and advantages emerge in these systems which cannot be found in their Hermitian counterparts \cite{zhang_pra_2012,lee_prl_2014}. Furthermore, the eigenspectrum makes the transition from real to imaginary around an exceptional point (EP) \cite{bender_prl_1998,bender_njp_2012} which has already been observed in optical setups, cold atoms and cavity systems \cite{kreibich_pra_2014,chitsazi_prl_2017,miri_science_2019,pan_pra_2019} and also plays a crucial role in the study of open system dynamics, especially, for the case of thermal machines \cite{khandelwal_prx_2021}.

In similarity, a quantum phase transition, separating two quantum phases of a Hermitian many-body systems, signifies the sudden change of properties at zero temperature driven by quantum fluctuations~\cite{sachdev_2011}. It has also been discovered that during evolution, the behavior of the Loschmidt echo (the overlap of the initial and the evolved state) may discriminate between two scenarios: in one, the quenching and the initial Hamiltonian are in separate phases, while in the other, they are in the same phase \cite{heyl_prl_2013, Heyl_review_2018}. The phenomenon is referred to as  dynamical quantum phase transition (DQPT) \cite{heyl_prl_2015, karrasch_prb_2017,porta_sr_2020} (cf. \cite{Krish2004, Aditi2005, Krish2008}) which was  initially demonstrated  using the quantum transverse Ising model. In this regard, several counter-intuitive results are also reported \cite{Sirker14, Vajna14, bhattacharyya_SR_2015, yang_prb_2017, gurarie_pra_2019, jafri_sr_2019_9,jafri_prb_2019_10, stav_prb_2020,jafri_prb_2020_5, jafri_iop_2020_7, guan_prr_2021,modak_prb_2021, jafri_pra_2021_6,jafri_prb_2021_4,haldar_prx_2021, nandi_prl_2022, jafri_pra_2022_3, jafri_prb_2022_2, jafri_prb_2022_1}  which include nonuniformly spaced critical time in the  $XY$ model with uniform and alternate magnetic fields, detection of  phase boundary at finite temperature via
long-time average of the Loschmidt echo   \cite{nandi_prl_2022}, enhanced sensitivity in connection with quantum sensing \cite{guan_prr_2021}, the dynamical signature to disclose the localization-delocalization transition in the Aubry-Andr{\'e} (AA) model \cite{yang_prb_2017}. Since there are only a few many-body Hamiltonian which can be diagonalized analytically, the observation of DQPT can also be a potential method to predict a possible QPT at equilibrium. Moreover, DQPT has already been captured in laboratories with physical systems like cold atoms and trapped ions \cite{jurcevic_prl_2017,flaschner_nature_2018}. Notice that in all these investigations, the initial state is the ground state of a Hermitian Hamiltonian and the evolution is unitary.

In a similar fashion, the localization regime in the dynamics of the non-Hermitian Aubry–Andr{\'e} model with $\mathcal{PT}$-symmetry using long-time survival probability \cite{zhihao_pra_2021} and the effect of topology after quenching in the non-Hermitian gapless phase \cite{zhou_pra_2018, zhou_njp_2021, hamazaki_nc_2021, mondal_prb_2022, mondal_prb_2023, jafri_prb_2020_8} have been studied to illustrate the role of non-Hermiticity in spin models. Also, recent reports in non-Hermitian systems have shown the spreading of correlation and the change of block entanglement from volume law to area law \cite{turkeshi_prb_2023, gal_scipost_2022}. Further, the dynamics in the non-Hermitian systems under sudden quench has been investigated in optical systems \cite{wiersig_pra_2008}, and also in quantum systems like the Bose-Hubbard model \cite{Graefe_2008, graefe_pra_2010} and hydrogen atom \cite{cartarius_pra_2011}. Note that in all these previous works,  the initial state is prepared in the ground state of the Hermitian Hamiltonian while the system is evolved according to a non-Hermitian Hamiltonian. In contrast, in our work, the ground state is chosen as the initial state of the non-Hermitian spin model and the evolution occurs by abruptly changing the corresponding parameter of
the NH Hamiltonian.

Specifically, we explore the scenario where the ground state of NH Hamiltonian in the unbroken phase evolves according to a pseudo-Hermitian rotation-time reversal (\(\mathcal{RT}\))-symmetric model. The prominent examples of the \(\mathcal{RT}\)-symmetric Hamiltonian include the \(XY\) model with imaginary anisotropy parameter in the presence of a uniform and alternating transverse magnetic field, referred to as \(iXY\) and \(iATXY\) models respectively. Note that similar to Hermitian models \cite{barouch_pra_1970, barouch_pra_1971, divakaran_prb_2008, titas_pra_2016}, both models can be mapped to  spinless fermions \cite{Song_RT_symm, ganesh_aditi_factorization_surface}, making their dynamics analytically tractable. Starting from the ground state of the unbroken phase and employing the biorthogonalization technique, we demonstrate that the rate function derived from Loschmidt echo along with the averaged value of the Loschmidt echo in a short- and long-time can determine whether the quenching is performed across the EP or not. In particular, from the unbroken to broken quench, the rate function
increases monotonically and saturates to a nonvanishing value while the sudden quench to the unbroken phase, the rate function oscillates at a very low value. We illustrate that the nonanalytic behavior in the derivative of the long-time averaged rate function correctly indicates the exceptional points known for both the \(iXY\) and the \(iATXY\) models.

In addition to the exactly solvable models,  we deal with the \(iXYZ\) model having short- and long-range interactions which can be studied via exact diagonalization. In these non-Hermitian models also, we reveal that dynamical quantifiers,  both Loschmidt echo and its time average in the transient regime, can accurately locate the exceptional points,
thereby mimicking the study of long-time dynamics. It was noticed that finding the spectrum in non-Hermitian systems is relatively hard, thereby pinpointing EPs via exact diagonalization can face certain numerical problems \cite{bender_njp_2012} and should be solved with the $ARNOLDI$ algorithm \cite{arnoldi}.  Hence our results indicate that dynamical quantifiers, which are accessible in the current experimental setups can also be proposed as appropriate indicators for determining EPs.

This paper is organized as follows. In Sec. \ref{sec:dqpt}, we introduce the quantities that we use to analyze DQPT for short as well as long time, while we describe different non-Hermitian Hamiltonian in Sec. \ref{sec:hamilonianns}. Analytical computation of Loschmidt echo and the behavior of the rate function for the \(iXY\) and \(iATXY\) models which detect EP are presented in Sec. \ref{Sec:Loshcmidt_rate}. In Sec. \ref{sec:finding_ep}, we extend the scenario  to capture the exceptional point through transient dynamics for a system  like  \(iXYZ\) with short- and long-range interaction which can only be solved numerically. Finally, we draw conclusion of our findings in Sec. \ref{sec:conclusion}.

\section{Dynamical Quantifiers}
\label{sec:dqpt}

Dynamical quantum phase transitions have been quantified by the non-analytic behavior of a distance function of the evolved and the initial states \cite{Jozsa_axioms}, called the Loschmidt echo \cite{Heyl_review_2018,silva_l_echo_1, peres_l_echo_2}. It is defined as 
\begin{equation}
    \mathcal{L}(t)=\frac{\abs{\langle\Psi(0)|\Psi(t)\rangle}^2}{\langle \Psi(t)|\Psi(t)\rangle},
    \label{LE_def}
\end{equation}
where $|\Psi(0)\rangle $ 
 is the initial state of the Hamiltonian while $|\Psi(t)\rangle$ represents the evolved state after a quench.
 Typically, the initial state is considered to be the ground state of the Hamiltonian, $H(\lambda)$ at \(t=0\), corresponding to a certain phase while the evolution occurs due to a sudden change of parameters in the Hamiltonian at \(t>0\),  $H(\lambda^\prime)$ which may or may not belong to the same phase as $H(\lambda)$. We are interested in the situation when $|\Psi(t)\rangle $ is orthogonal to $|\Psi(0)\rangle$, thereby arriving at vanishing $\mathcal{L}(t)$. More vividly, one can capture this feature by determining the Loschmidt rate, which is defined as 
\begin{equation}
    \lambda(t)=-\lim_{N\to\infty}\frac{1}{N}\ln \mathcal{L}(t)
    \label{rate_def}
\end{equation}
where $N$ is the system size, and $N\to\infty$ is taken for the  thermodynamic limit. Note that the rate function is an analog of free energy as it is the logarithm of a function that resembles the partition function when the corresponding inverse temperature is replaced with the imaginary time. The nonanalyticities of the rate function, known as Fisher zeros, indicate the crossing of a phase boundary and hence, equilibrium phase transition can be replicated using nonequilibrium dynamics successfully in Hermitian systems \cite{heyl_prl_2013} (for exceptional cases, see \cite{Vajna14, gurarie_pra_2019, stav_prb_2020, nandi_prl_2022}).

In this work we will employ these quantifiers, both \(\mathcal{L}(t)\) and \(\lambda(t)\), to study non-Hermitian systems. Specifically, instead of quantum critical points, our aim is to detect the exceptional point which distinguishes between broken and unbroken phases in the non-Hermitian systems from the properties of the evolved state. Precisely, the initial state is chosen from the unbroken phase (the entire spectrum of the Hamiltonian is real), and then the system is quenched to either the broken (some eigenvalues have nonzero imaginary component)  or the unbroken phase. We investigate $\mathcal{L}(t)$ or $\lambda(t)$ to predict whether the system has crossed the exceptional point or not. In the case of non-Hermitian models, eigenvectors are not, in general, orthogonal which can be made orthogonal by defining an inner product with respect to a metric, known as biorthogonalization (see Appendix \ref{appendix:biortho}). This process ensures that the initial Hamiltonian $H(\lambda)$ has orthogonal eigenvectors. As mentioned before, although in many situations the Loschmidt rate can identify equilibrium phase transition successfully, there are Hamiltonian including the quantum $XY$ model with uniform and alternating magnetic fields for which the quantifiers fail \cite{Vajna14, gurarie_pra_2019, stav_prb_2020, nandi_prl_2022}. To avoid such discrepancy, in the case of Hermitian systems, time averaged Loschmidt rate and Loschmidt echo are used \cite{eta_paper}. We choose the same path since the non-Hermitian models considered in this work resemble the Hermitian $XY$ model.

To determine exceptional points through dynamics for a non-Hermitian Hamiltonian, we adopt four figures of merits, time average of the Loschmidt echo in the transient (steady) state regimes, and the corresponding rate functions.
Mathematically,  the rate functions  are defined as
\begin{eqnarray}
\eta^T &=&-\frac{1}{N}\ln(\frac{1}{\tau_0}\int_{0}^{\tau_0}\mathcal{L}(t)\dd t),\\
   \mbox{and} \, \,  \eta^S &=&-\frac{1}{N}\ln(\lim_{\tau\to\infty}\frac{1}{\tau-\tau_1}\int_{\tau_1}^{\tau}\mathcal{L}(t)\dd t),\nonumber
   \label{eq:eta_s}
\end{eqnarray}
where $\tau_0$ and $\tau_1$ describe the end of transient time and beginning time of the steady state regime respectively. It has been shown  \cite{eta_paper} that $\eta^S$ can be used to distinguish between phases, thereby confirming the existence of quantum critical points in different Hermitian models. We propose here that $\eta^S$ (averaged Loschmidt echo) is capable of differentiating broken and unbroken phases in the steady state domain while $\eta^T$ is the similar quantity calculated in the transient regime. It turns out that the latter quantity is better for models that can only be solved numerically since the finite size effects accumulate in the long-time dynamics while both quantities are effective for models like $iXY$ and $iATXY$ models which can be solved analytically.

\section{$\mathcal{RT}$-symmetric spin models}
\label{sec:hamilonianns}

Non-Hermiticity in spin models can be introduced either by imaginary magnetic fields \cite{giorgi_prb_2010} or by considering imaginary anisotropy parameter in the interactions \cite{Song_RT_symm} that can reveal several novel features \cite{bender_ropp_2007}. In this regard, rotational-time ($\mathcal{RT}$) symmetry was found to be one of the facets through which the spectrum can be made real in a non-Hermitian interacting spin systems.

The rotation operator rotates each spin by angle $\frac{\pi}{2}$ about the $z$-axis, i.e., $\mathcal{R}=\exp[-i\frac{\pi}{4}\sum_{i}^{N}\sigma_i^z]$ while the time reversal operator, \(\mathcal{T}\), which is the complex conjugation in finite-dimensional system is an anti-linear operator, i.e.,  $\mathcal{T}i\mathcal{T} = -i$. Although Hamiltonian commutes with the $\mathcal{RT}$ operator (it does not commute with  $\mathcal{R}$ and  $\mathcal{T}$ operators individually), they may not share the common eigenvectors due to the anti-linearity of the time reversal operator $\mathcal{T}$, where the spectrum becomes
complex, known as the broken phase. However,  when they do share the eigenspectrum, the Hamiltonian has the real eigenvalues and is in the unbroken phase. Hence, by varying parameters of the Hamiltonian, we can traverse between the broken and the unbroken regions.

\subsection{Non-Hermitian $iXY$ Hamiltonian}
\label{NH_iXY_Hamil}

The rotation operator $\mathcal{R}$ essentially maps Pauli matrix $\sigma^x$ to $\sigma^y$ and vice versa while  $\sigma^z$ remains unchanged as the rotation is about the same. Thus, to look for the models that are endowed with $\mathcal{RT}$-symmetry, we should look for the cases that have complex conjugated coefficients from the interactions or the magnetic fields in the $x$ and $y$ directions.

We investigate a variety of possible models having \(\mathcal{RT}\)-symmetry. Replacing the anisotropy parameter, $\gamma$ by $i\gamma$ in the prototypical quantum $XY$ model with nearest neighbor interactions in a transverse magnetic field, we obtain the $iXY$ model which possesses $\mathcal{RT}$-symmetry, 
\begin{equation}
    \bar{H}_{iXY} = \sum_{l=1}^{N}\frac{J}{4}\left [(1+i\gamma)\sigma^x_l\sigma^x_{l+1}+(1-i\gamma)\sigma^y_l\sigma^y_{l+1}\right ] - \frac{h^\prime}{2}\sigma^z_l,
    \label{iXYRT}
\end{equation}
with the periodic boundary condition, $\sigma_{N+1}^i\equiv \sigma_1^i$ ($\{ i\in x,y,z\}$), $J$ and $h^\prime$ being the coupling constant and the strength of magnetic field respectively. We set $\frac{h^\prime}{J} = h$.

Applying the same recipe of the Hermitian $XY$ models \cite{barouch_pra_1970, barouch_pra_1971},  i.e., after the Jordan-Wigner and Fourier transformations (see Appendix \ref{appendix:diag}),  the block-diagonal form for each mode in the momentum phase represented as  $\bar{H}^p_{iXY} = \hat{A}_p^\dagger H^p_{iXY} \hat{A}_p,$ where \(\hat{A}_p\) is the column vector, \((\hat{a}_p, a_{-p}^\dagger)^T\), 
\begin{equation}
    H^p_{iXY} = 
   \left[ \begin{array}{cc}
h -\cos\phi_p & -\gamma \sin \phi_p \\
\gamma \sin \phi_p & \cos \phi_p - h
\end{array}\right]
\end{equation} and \(\phi_p=\frac{2\pi p}{N}\) with \(p\in[1, \frac{N}{2}]\) with $\hat{a}_p^\dagger$ and $\hat{a}_p$ being the fermionic creation and annihilation operators in the momentum space respectively. In the thermodynamic limit, i.e., as $N \rightarrow \infty$, we land in the scenario with $\phi_p \in (0,\pi]$. Therefore, the eigenvalues of $\bar{H}^p_{iXY}$ are
\begin{equation}   
   \epsilon^\mp_p= \mp\sqrt{(h-\cos\phi_p)^2-\gamma^2\sin^2\phi_p},
   \label{eq: HiXYRT_eval}
\end{equation}
where the corresponding eigenvectors are $|v^-_p\rangle$ and $|v^+_p\rangle$ respectively. They are written in the columns of a matrix, $P_p$, as 
\begin{eqnarray}
    P_p = 
    \left[\begin{array}{cc} \text{\footnotesize $\sqrt{h-\cos\phi_p-\epsilon_p}$} & \text{\footnotesize $-\sqrt{h-\cos\phi_p+\epsilon_p}$}\\ \text{\footnotesize $-\sqrt{h-\cos\phi_p+\epsilon_p}$} & \text{\footnotesize $\sqrt{h-\cos\phi_p-\epsilon_p}$} \end{array}\right],  
    \label{eq: HiXYRT_evec}
\end{eqnarray}
where $\epsilon_p = |\epsilon^\mp_p|$. Note that $|v^-_p\rangle$ and $\epsilon^-_p$ contribute to the ground state and ground state energy respectively for each momentum $\phi_p$.

\subsection{Non-Hermitian \(iATXY\) model}

Another prominent quantum spin model is the nearest-neighbor $XY$ model with alternating magnetic fields, i.e., the $ATXY$ model, which was shown to possess a richer phase diagram than the $XY$ model \cite{divakaran_prb_2008, titas_pra_2016}. Its non-Hermitian counterpart, referred to as the $iATXY$ model can also be analytically solved \cite{ganesh_aditi_factorization_surface}. 
The governing Hamiltonian reads as
\begin{equation}
    \bar{H}_{iATXY} =  \bar{H}_{iXY}+\sum_{l = 1}^N\frac{(-1)^l h_a^\prime}{2}\sigma^z_l,\nonumber
\end{equation}
with $\frac{h^\prime \pm h_a^\prime}{J} =h\pm h_a $  being the strength of the magnetic field at even and odd sites respectively.

Unlike the $iXY$ model, the modified Jordan-Wigner transformation \cite{Saptarshi_titas_atxy, ganesh_aditi_factorization_surface} changes the one-dimensional model to two-component Fermi gas based on even and odd sites, followed by the Fourier transform. The final $4\times 4$ momentum block \cite{ganesh_aditi_factorization_surface, divakaran_prb_2008} can be written in the basis of $S_p =\{a_p^\dagger, a_{-p}, b_p^\dagger, b_{-p}\}$ as $\bar{H}_{iATXY} = \bigoplus_{p} S_p H^p_{iATXY} S_p^\dagger$ such that 
\begin{align}
H^p_{iATXY} &= \nonumber \\ & \left[\begin{array}{cccc}
h+\cos\phi_p & -\gamma \sin\phi_p & 0 & -h_a \\
\gamma \sin\phi_p & -h-\cos\phi_p & h_a & 0 \\
0 & h_a & \cos\phi_p-h & -\gamma \sin\phi_p \\
-h_a & 0 & \gamma \sin\phi_p & h-\cos\phi_p
\end{array}\right], 
\label{eq:iatxy_mom_block}
\end{align} 
where $a_p $ and $b_p$ are fermionic operators that correspond to even and odd sites respectively, with $\phi_p\in [-\pi/2,\pi/2]$ in the thermodynamic limit. The eigenvalues of the momentum block are given by $\pm\tilde{\epsilon}_p^{\pm}$ where
 \begin{equation}
\begin{aligned}
    \tilde{\epsilon}_p^{\pm} = &\Big(h^2+h_a^2+\cos^2\phi_p-\gamma^2\sin^2\phi_p\\
    &\left.\pm 2\sqrt{h^2 h_a^2+h^2\cos^2\phi_p-h_a^2\gamma^2\sin^2\phi_p}\right )^{1/2}.
\end{aligned}
\end{equation}
Note that calculations with the exact form of the eigenvectors will be too cumbersome, rather they are computed following numerical diagonalization of $H^p_{iATXY}$, which are used for biorthogonalization (see Appendix \ref{appendix:biortho}) during the dynamics.

\subsection{Non-Hermitian $iXYZ$ models: Short- and Long-range interactions}

Up to now, all the non-Hermitian models that we discussed are analytically diagonalizable. Apart from these models, we also consider a model which can be solved only by numerical means or approximate methods. Specifically, we deal with $iXYZ$ models having $\mathcal{RT}$-symmetry with short- $(SR)$ and long-range $(LR)$ interactions, given, respectively by
\begin{eqnarray}
    \nonumber\bar{H}_{iXYZ} &=& \sum_{l=1}^{N}\frac{J}{4}\left [(1+i\gamma)\sigma^x_l\sigma^x_{l+1}+(1-i\gamma)\sigma^y_l\sigma^y_{l+1}\right.\\&&\left.+\Delta\sigma^z_l\sigma^z_{l+1}\right ] - \sum_{l=1}^{N}\frac{h^\prime}{2}\sigma^z_l,
    \label{eq:iXYZRT}
\end{eqnarray}
and
\begin{eqnarray}
    \nonumber\bar{H}^{LR}_{iXYZ} &=& \sum_{l,m}\frac{J_{lm}}{4}\left [(1+i\gamma)\sigma^x_l\sigma^x_m+(1-i\gamma)\sigma^y_l\sigma^y_m\right.\\&&\left. +\Delta\sigma^z_l\sigma^z_m\right ] - \sum_{l=1}^{N}\frac{h^\prime}{2}\sigma^z_l,
    \label{eq: iXYZRT_LR}
\end{eqnarray}
with periodic boundary condition,  $\Delta$ being the strength of interactions in the $z$-direction, $J_{lm} = \frac{J}{|l-m|^\alpha}$ with $\alpha$ being the falling-rate of the interactions with distance between the spins, and $\frac{h^\prime}{J} = h$ being the strength of the magnetic field. In the Hermitian cases, the former model can be realized in photonic systems \cite{ma_np_2011} while the latter models are found in experiments with trapped ions \cite{bermudez_prb_2017} and superconducting circuits. Moreover, the long-range Hermitian models typically reveal novel characteristics like violation of area-law \cite{nielsen_pra_2011,gong_prl_2017}, and fast transfer of states \cite{eldredge_prl_2017}, which cannot be found in models having $SR$ interactions.

\begin{figure}
    \includegraphics[width=\linewidth]{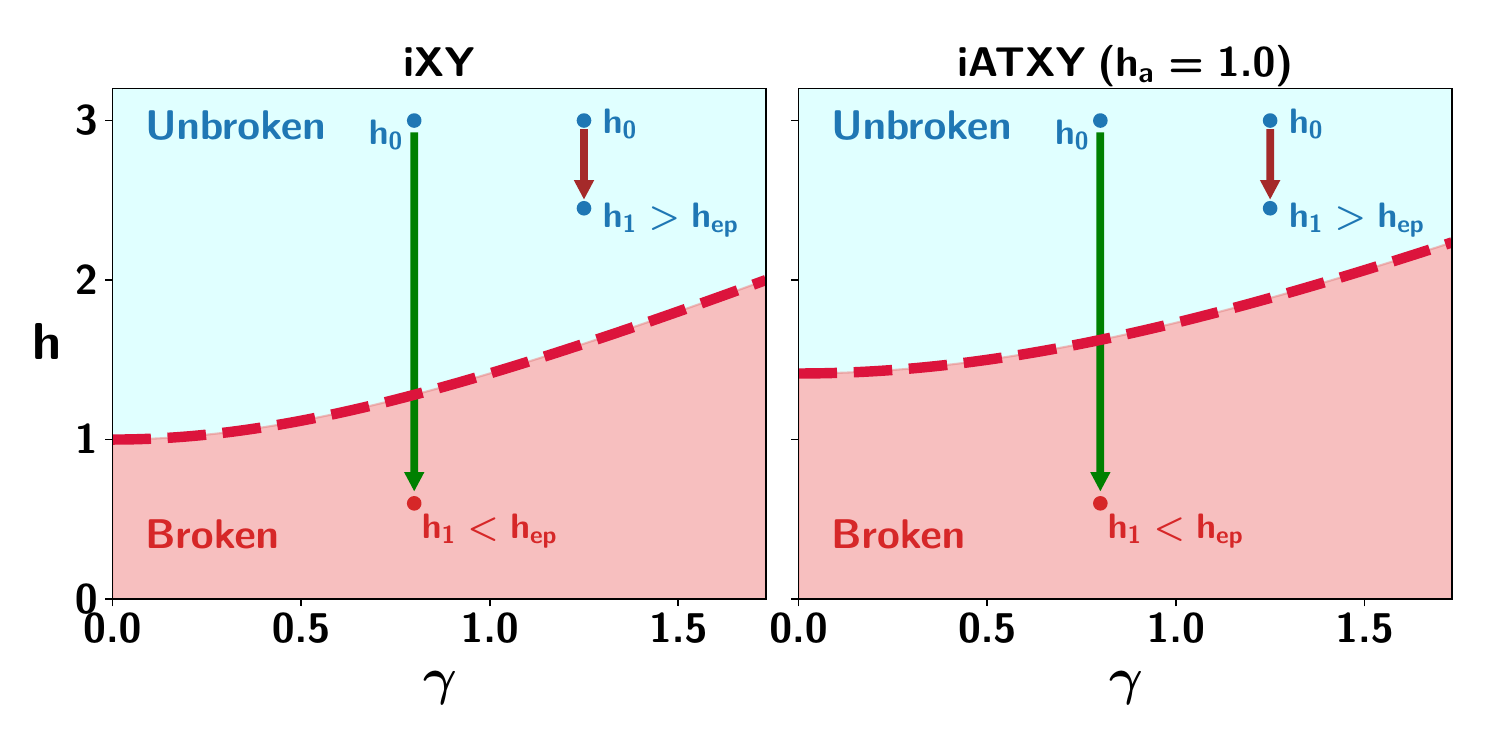}
    \caption{ Parameters for the $iXY$ and the $iATXY$ model. The blue (light) shaded region represents the unbroken phase, and the broken phase is marked as red (darker), and the red dashed line represents the exceptional points. Starting with the ground state at magnetic field strength $h_0$, different quenching scenarios are depicted when the magnetic field strength is changed to $h_1<h_{ep}$ or $h_1>h_{ep}$ with \(h_{ep}\) being the exceptional point.}
    \label{fig:param}
\end{figure}

\section{Distinguishing exceptional point via Loschmidt Rate in analytical models}
\label{Sec:Loshcmidt_rate}
A defining property of the non-Hermitian Hamiltonian is the existence of exceptional points. This is the point that distinguishes two regions having real and complex energy spectra, corresponding to the unbroken and the broken phases respectively. We first discuss the procedure to find and calculate the exceptional points of all the models considered in this work.  We then present our results on detecting exceptional points using the dynamical quantifiers described in the previous section.

\subsection{Exceptional points}
The exceptional points of the models discussed in the previous section have been studied in the literature. The $iXY$ and the $iATXY$ models are analytically solvable and the exceptional points are exactly known \cite{Song_RT_symm, ganesh_aditi_factorization_surface}. After finding the energy spectrum of the $iXY$ and the $iATXY$ models, one needs to solve for two equations, namely, $\frac{\partial \epsilon_p (h_{ep})}{\partial p } = 0$ and $\epsilon_p(h_{ep}) = 0$ to find the momentum value $p$ at which the energy vanishes   with \(\epsilon_p\) being the energy spectrum in terms of momentum basis and $h_{ep}$ being the exceptional point. An exceptional point can be obtained by solving these two equations simultaneously which collectively imply that when the ground state energy becomes zero, the system undergoes a transition from a broken  to an unbroken phase.

The unbroken and broken phases are classified based on the symmetry followed by the ground state eigenvector. For a given Hamiltonian $H$ following $\mathcal{RT}$ symmetry, i.e., $\left[H, \mathcal{RT}\right] = 0 $, and when the ground state has the $\mathcal{RT}$ symmetry, $H$ is in the unbroken phase while $H$ is in the broken phase when the ground state breaks the $\mathcal{RT}$ symmetry. Using the above procedure, one finds that for both the $iXY$ and the $iATXY$ models, we have $h>h_{ep}$ in the unbroken phase (nondefective), while $h\leq h_{ep}$ is in the broken phase (defective) (see Fig. \ref{fig:param} for schematic representation and Appendix \ref{appendix:defective} for the detailed discussions on defective and nondefective Hamiltonian), with 
\begin{equation}
    h_{ep}^{iXY}=\sqrt{1+\gamma^2} \text{ and } h_{ep}^{iATXY}=\sqrt{1+h_a^2+\gamma^2}.
    \label{eq:excep_XY_atxy}
\end{equation}  

It has recently been shown by some of us that the exceptional points in the $\mathcal{RT}$-symmetric Hamiltonian are closely related to the factorization surface where the ground state of the  Hamiltonian is fully separable with vanishing entanglement of the corresponding Hermitian model 
\cite{ganesh_aditi_factorization_surface}. This prescription allows one to find the exceptional point even for the systems that have no analytical solution. For instance, the above connection leads to the exceptional point as \begin{equation}
    h^{iXYZ}_{ep} = \sqrt{(1+\Delta)^2+\gamma^2}
    \label{eq:excep_XYZ}
\end{equation} for the $iXYZ$ model in Eq. (\ref{eq:iXYZRT}) which does not have an analytical solution and can be confirmed via numerical diagonalization. This procedure also suggests that in the presence of long-range interactions of the form of Eq. (\ref{eq: iXYZRT_LR}), the exceptional point satisfies the equation, given by 
\begin{equation}
    h^{iXYZ_{LR}}_{ep}= \sqrt{(1 + \Delta)^2 + \gamma^2} \sum_{j=1}^{N/2}\frac{1}{j^\alpha}.
    \label{eq:excep_LR}
\end{equation}

\begin{figure*}
    \includegraphics[width=\linewidth]{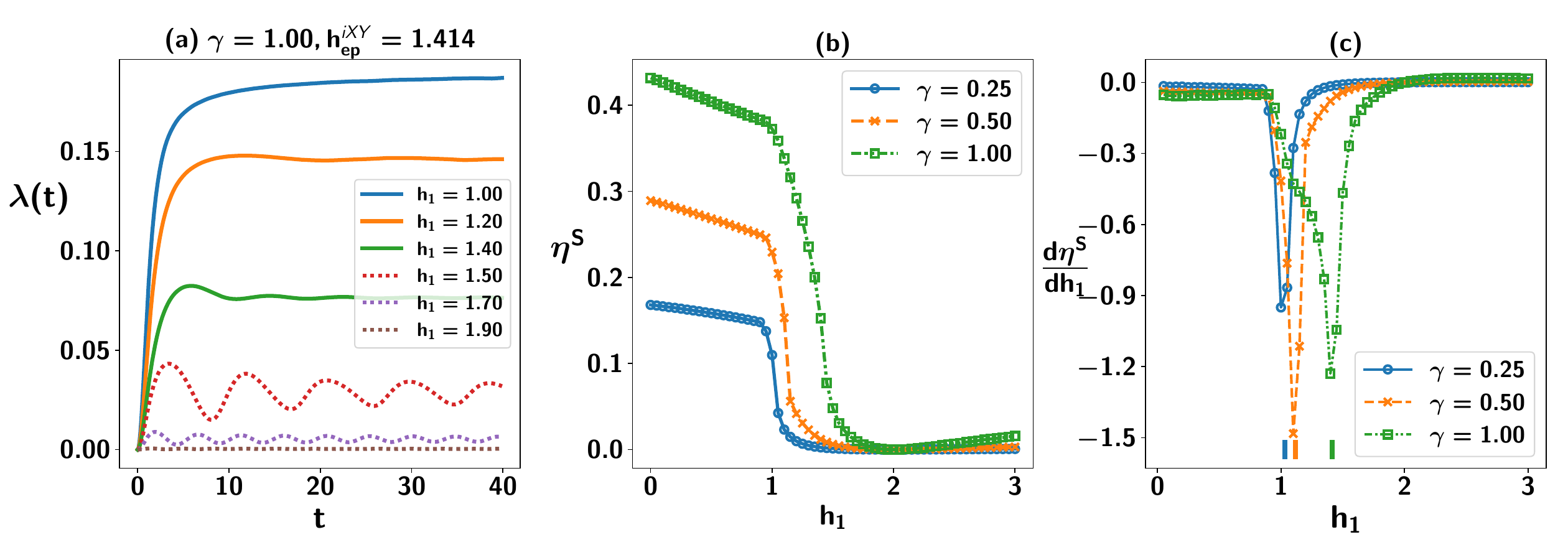}  
    \caption{
    Quench for $iXY$ model with the initial state as the ground state at $h_0=2.0>h^{iXY}_{ep}$ for all $\gamma$ considered. (a) Rate function $\lambda(t)$ (ordinate) is plotted with respect to time (\(t\)) (abscissa) for the $iXY$ Hamiltonian in Eq. (\ref{iXYRT}). The sudden quench is performed by changing the magnetic fields to \(h_1\) given in legends. Solid lines represent quenching from unbroken to broken phase, while dotted lines are for the quenching in the same phase.  
    (b) Distinguishing between the broken and unbroken phases using $\eta^S$ (vertical axis) defined in Eq. (\ref{eq:eta_s}) against \(h_1\) (horizontal axis) for different values of \(\gamma\).  For calculating $\eta^S$,  the averaging is performed during $t \in [20, 200] $, i.e. $\tau_1=20$.
    (c) Derivative of $\eta^S$ with respect to \(h_1\) (vertical axis) is plotted with post-quenched magnetic fields, (\(h_1\)) (horizontal axis). The nonanalyticity of \(\frac{d\eta^S}{d h_1}\) signals the EP marked underneath by a vertical bar which can be obtained analytically with $N=1200$ (see Eq. (\ref{eq:excep_XY_atxy})). All the axes are dimensionless.}
    \label{fig:xy_rt}
\end{figure*}

\subsection{Dynamics of non-Hermitian models }

In order to present our results for the \(iXY\) model, let us first fix the stage.  After the preparation of the initial state corresponding to the $iXY$ Hamiltonian at magnetic field strength $h_0$, whose momentum blocks are denoted by \((H^0)^p_{iXY}\), the system is evolved by suddenly changing the external magnetic field, \(h\) from $h_0$ to $h_1$, whose momentum blocks are denoted by $(H^1)^p_{iXY}$. As mentioned before, the parameters involved in the Hamiltonian for the initial state are fixed in the unbroken phase while the Hamiltonian corresponding to the evolution operator is chosen from both the broken and the unbroken phases, thereby producing the dynamical state which is the main interest of the paper. 
Specifically,  the initial magnetic field is adopted such that $h_0 > h_{ep}^{iXY}$,
while at \(t>0\), the quench  is performed by changing the external magnetic field to \(h_1\). Notice that the corresponding eigenvalues are given by $ \epsilon_p^{0,1}= \sqrt{(h_{0,1}-\cos\phi_p)^2-\gamma^2\sin^2\phi_p}$, where $\pm\epsilon^0_p$ and $\pm\epsilon^1_p$ are the eigenvalues obtained from the initial and the quenched Hamiltonian momentum blocks respectively. In the eigenspace of $(H^0)^p_{iXY}$ (given in Sec. \ref{NH_iXY_Hamil}) with the magnetic field strength $h_0$, which is, equivalently, bi-orthogonalizing the vectors, the initial Hamiltonian would be diagonal, i.e.,
  $ (H^0)^p_{iXY} = \left[\begin{array}{cc}
-\epsilon_p^0 & 0 \\
0 & \epsilon_p^0
\end{array}\right]$,
and the final Hamiltonian,   with which we perform the quench, written in the same basis as of \((H^0)^p_{iXY}\), reads as 
\begin{equation}
    (H^1)^p_{iXY} = \frac{1}{\epsilon_p^0}\left[\begin{array}{cc}
-\delta_p & \omega_p \\
-\omega_p & \delta_p
\end{array}\right],
\end{equation}
where $\delta_p = (h_0-\cos\phi_p)(h_1-\cos\phi_p) - \gamma^2\sin^2\phi_p$ and $\omega_p = \gamma(h_1-h_0)\sin\phi_p$.  
Note that  $(H^1)^p_{iXY}$ is still non-Hermitian. 
Therefore, the time-evolution operator, in this case, can be represented as
\cite{zhou_pra_2018}, 
\begin{equation}
    U^p_{iXY}(t) = e^{-i (H^1)^p_{iXY} t} = \cos\epsilon_p^1 t\:\mathbb{I} - i \frac{\sin\epsilon_p^1 t}{\epsilon_p^1} (H^1)^p_{iXY}.
\end{equation} 
Since the initial Hamiltonian is diagonal,  its ground state is $|\psi_p(0)\rangle = \left[1 \text{  } 0\right]^T$ and the evolved state can take the form as 

\begin{eqnarray}
   |\psi_p(t)\rangle &=& U^p_{iXY}(t)|\psi_p(0)\rangle\nonumber\\
   &=& \left[\begin{array}{c}
       \cos\epsilon_p^1 t - i \frac{\delta_p\sin\epsilon_p^1 t}{\epsilon_p^0 \epsilon_p^1} \\
       i \frac{\omega_p\sin\epsilon_p^1 t}{\epsilon_p^0 \epsilon_p^1}
       \end{array}\right] \coloneqq \left[\begin{array}{c}
       u_p(t) \\ v_p(t)
       \end{array}\right].
\end{eqnarray}
We normalize the dynamical state due to the non-Hermiticity of the quenching Hamiltonian. Finally, for the $iXY$ model, the expression for the Loschmidt echo reads as \begin{equation}
     \mathcal{L}(t) = \prod_p\mathcal{L}_p(h_0,h_1,\gamma,t) = \prod_p\left (\frac{|u_p(t)|^2}{|u_p(t)|^2+|v_p(t)|^2} \right ).
 \end{equation}

We now show that by analyzing $\mathcal{L}(t)$, one can distinguish between broken and unbroken phases, thereby identifying exceptional points. A general matrix representation of a non-Hermitian Hamiltonian $H$ can be written in Jordan canonical form $\mathcal{J}$, with respect to some basis (columns of matrix $\mathcal{S}$), such that $H = \mathcal{S} \mathcal{J} \mathcal{S}^{-1}$. These are decomposed in Jordan blocks, $\mathcal{J} = \bigoplus_l \mathcal{J}_l$ with complex or real eigenvalues $\theta_l$. When the order of degeneracy in the eigenvalue  is equal to the eigenspace dimension, Jordan blocks are diagonal and give the eigenvectors in the unbroken phase, where the Hamiltonian is \textit{nondefective}. The evolved state by the action of the evolution operator $U(t) = \bigoplus_l U_l(t)$ on $|\Psi(0)\rangle$ is given by
\begin{equation}
    |\Psi(t)\rangle = U(t)|\Psi(0)\rangle = \mathcal{S} e^{-i\mathcal{J} t} \mathcal{S}^{-1}|\Psi(0)\rangle.
\end{equation}    
On one hand, in the unbroken phase, the Jordan form of each $(H^1)^p_{iXY}$ is diagonal with real eigenvalues $\mp\epsilon^1_p$ and complete eigenbasis, $\mathcal{S}_p$. Therefore, in the eigenbasis $\mathcal{S}_p$, the initial state reads as $|\psi_p(0)\rangle = c_1|\mathcal{S}^p_1\rangle +c_2|\mathcal{S}^p_2\rangle$, where $|\mathcal{S}^p_{1,2}\rangle$ are the first and the second columns of $\mathcal{S}^p$ respectively, and $\langle\psi_p(0)| = c_1^*\langle S^{-1}_1| +c_2^*\langle S^{-1}_2|$ where $\langle \mathcal{S}^{-1}_{1,2}|$, represent the first and the second rows of $(\mathcal{S}^{p})^{-1}$, with $|c_1|^2+|c_2|^2=1$. After quenching in the unbroken phase, the Jordan form is diagonal and the Loschimdt echo reduces to
 \begin{eqnarray}
\mathcal{L}^\mathcal{U}_p(t) &=& |\;|c_1|^2e^{i\epsilon^1_p t}+|c_2|^2e^{-i\epsilon^1_p t}\;|^2\\
\nonumber &=&\cos^2\epsilon^1_p t +(|c_1|^2-|c_2|^2)^2\sin^2\epsilon^1_p t,
\end{eqnarray}
where $\pm\epsilon^1_p$ are the real eigenvalues of quenching Hamiltonian. It indicates that  $\mathcal{L}^\mathcal{U}_p(t)$ is oscillatory for all $t$ and for all momentum $p$ in the unbroken phase. We will show that this is indeed the case [see Fig. \ref{fig:xy_rt}(a) when $h > h_{ep}^{iXY}$]. 

On the other hand, when we quench in the broken phase, the Hamiltonian becomes defective with the vanishing degenerate eigenvalue for certain critical momentum $\phi_{p^c}$. The defective $2\cross 2$ traceless Hamiltonian, $H_{p^c}$, in the Jordan canonical form $\mathcal{J}_{p^c}$ becomes
$\mathcal{J}_{p^c} = \left[\begin{array}{cc}
0 & 1 \\
0 & 0 
\end{array}\right]
\implies e^{-i\mathcal{J}_{p^c}t} = \left[\begin{array}{cc}
0 & -it \\
0 & 0
\end{array}\right].
$
The Loschmidt echo in this situation reduces to 
\begin{equation}
\mathcal{L}^\mathcal{B}_{p^c}(t) = \frac{|1-itc_2c_1^*|^2}{|c_1|^2+t^2|c_2|^2-itc_1^*c_2+itc_2^*c_1},
\end{equation}
 where both the numerator and the denominator are polynomial in $t$. With increasing $t$, the numerator is dominated by the term, $t^2|c_1|^2|c_2|^2$ and the denominator by $t^2|c_2|^2$. Therefore, $\lim_{t\to\infty}\mathcal{L}^\mathcal{B}_{p^c}(t)=|c_1|^2<1$ as the initial and the final Hamiltonian are noncommuting and do not share the basis of the Jordan form. It implies that at the beginning of the evolution, the Loschimdt echo decays if the quenching Hamiltonian is defective.  Such analysis demonstrates that the difference in the behavior of a dynamical quantity, namely, $\mathcal{L}(t)$, in two different phases, broken and unbroken, is capable of predicting the existence of exceptional points, without the explicit computation of eigenvalues.

In the case of the $iATXY$ model, exactly solving the Hamiltonian, $H^p_{iATXY}$ and finding the eigenvectors for arbitrary parameters is cumbersome. We numerically diagonalize $4\cross4$ block in Eq. (\ref{eq:iatxy_mom_block}) to obtain the corresponding metric and vectors in the momentum space which enable us to compute quantities for this model with a large system-size or in the thermodynamic limit. 
By numerically computing $\mathcal{L}_p(t)$, we  calculate the rate function $\lambda(t)$, defined in Sec. \ref{sec:dqpt}. For both $iXY$($h_a=0$) and $iATXY$ models, the rate function in the thermodynamic limit reduces to
\begin{equation}
    \lambda(h_0,h_1,\gamma,h_a,t) = -\int \frac{\dd\phi_p}{2\pi}\ln (\mathcal{L}_p(h_0,h_1,\gamma,h_a,t)),
    \label{rate_integral}
\end{equation}    
which is a function of all the parameters involved in the system.

\begin{figure*}
    \includegraphics[scale=0.5]{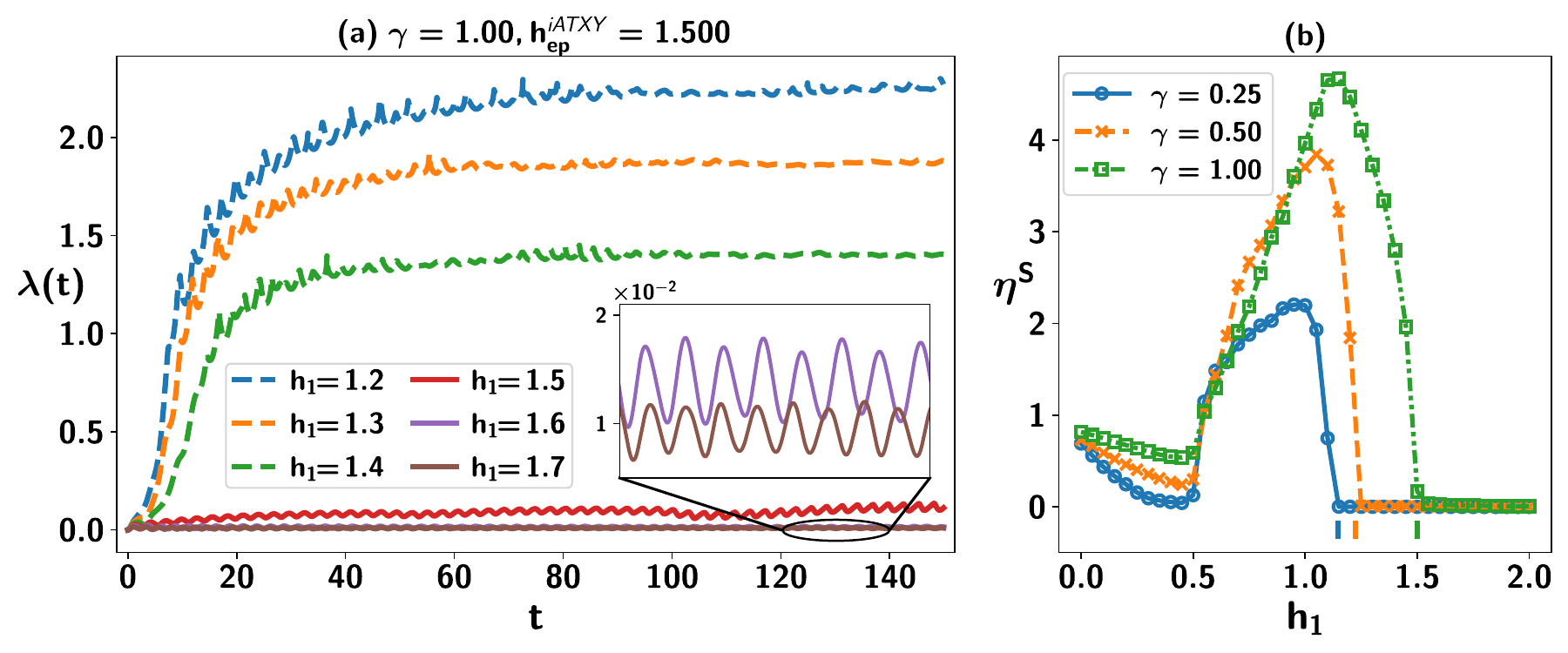}
    \caption{
     Quench for $iATXY$ model with the initial state as the ground state at $h_a=0.5, h_0=3.0>h_{ep}^{iATXY}$. (a) Rate function, $\lambda(t)$, (\(y\)-axis) as a function of \(t\) (\(x\)-axis). Dotted lines represent quenching from unbroken to broken phase, while solid lines are for the quenching in the same phase. (b) Rate of average Loschmidt echo ($\eta^S$) (ordinate) in the steady state regime  against the post-quench magnetic field (\(h_1\)) (abscissa). From the analysis, we find that the steady state is reached at $t=100$, which is used to compute $\eta^S$ during \(\tau_1=100\leq t \leq 500\) with $N=100$. The corresponding actual exceptional points are marked underneath by a vertical bar [see Eq. (\ref{eq:excep_XY_atxy})]. (Inset) A magnified view of the oscillations in the rate function with respect to \(t\) due to the quench in the unbroken phase. All the axes are dimensionless.}
    \label{fig:atxy_rt}
\end{figure*}

{\it Analysis of rate function and time averaged Loschmidt  echos. }
Let us now analyze the profile of  the rate function, $\lambda(t)$ or the Loschimdt echo, $\mathcal{L}(t)$ for different systems starting from the ground state of $H^0$. We mainly concentrate on two scenarios : (1) unbroken to broken quench and 
 (2) unbroken to unbroken quench. Notice that in the $iXY$ and the $iATXY$ models,  we look at the behavior of $\lambda(t)$, since  the Loschmidt echo $\mathcal{L}(t)$ becomes exponentially small with increasing system size $N$, thereby  exponentially increasing Hilbert space dimension.

{\it I. Unbroken to broken quench.} 
Since we are interested to locate the exceptional point of a non-Hermitian model, we first consider the situation where the initial state is prepared in the unbroken phase and the system is quenched to the broken phase of the model.

In the $iXY$ model, the rate function shows a steep rise in the transient regime ($0$ to $\tau_0=10$)  
[solid lines in  Fig. \ref{fig:xy_rt}(a) by taking the ground state at $h_0=2.0>h^{iXY}_{ep}$ as the initial state] when $h_1 < h_{ep}$ and saturates to a nonvanishing value near the exceptional point  in the  steady-state domain (from $\tau_1= 20>\tau_0$ to $\infty$).  In Fig.  \ref{fig:xy_rt}(a),  the anisotropy parameter is chosen to be unity, i.e., $\gamma=1.0$, with the corresponding exceptional point at $h^{iXY}_{ep}= \sqrt{2}$.
We perform the convergence test for the numerical integration of the rate function and find that $\lambda(t)$ converges at $N=1000$ in the $iXY$ model and hence we choose \(N=1200\) for illustration in Fig. \ref{fig:xy_rt}.

On the other hand, for a fixed \(\gamma\), $\eta^S$ decreases uniformly with increasing $h_1$ (by a step of $0.05$), till $h_1=1$, and then it decreases as a concave function, until the exceptional point as depicted in Fig. \ref{fig:xy_rt}(b). The change of curvature from concavity to convexity in $\eta^S$ leads to a sharp kink in $\frac{d\eta^S}{dh_1}$ at the exceptional point which is shown in Fig. \ref{fig:xy_rt}(c) by small vertical bars above the horizontal axis as obtained analytically. It clearly displays that the kink in dynamics is in good agreement with the exact values of the exceptional points obtained in the ground state using the Eq. (\ref{eq:excep_XY_atxy}) \cite{Song_RT_symm}. 

\begin{figure*}
    \includegraphics[scale=0.5]{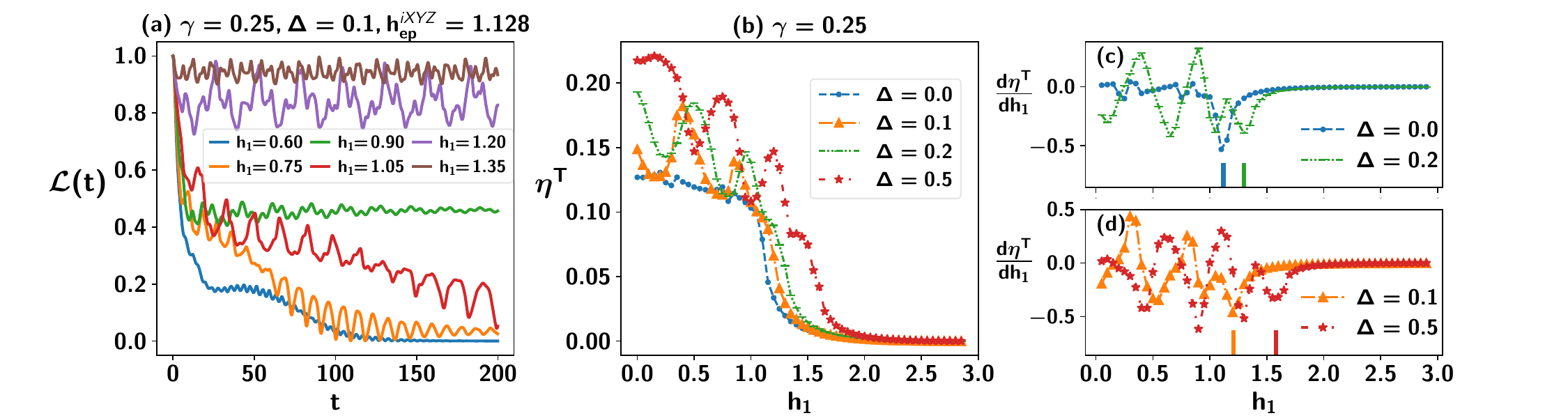}
    \caption{ Evolution of the nearest-neighbor $iXYZ$ model with the  magnetic field $h_0=3.0>h_{ep}^{iXYZ}$ in which the initial state is prepared, for different $\Delta$ and $\gamma$. (a) Loschidmt echo, $\mathcal{L}(t)$, (ordinate) with respect to time ($t$) (abscissa), for $\Delta=0.1$ and $\gamma=0.25$ and quenching in the corresponding broken and unbroken phases. The steady-state dynamics shows irregularity which can be due to finite size. (b) Short-time average of Loschmidt echo,  $\eta^T$ (ordinate), vs post quenched magnetic field, \(h_1\)(abscissa) for various $\gamma$ and $\Delta$.  The transient regime is taken to be $t=0$ to $\tau_0=30$. (c), (d) Behavior of $\frac{d\eta^T}{dh_1}$ (vertical axis) with respect to \(h_1\) (horizontal axis) for different $\Delta$ values. The corresponding exceptional points obtained analytically are marked by a vertical bar [see Eq.( \ref{eq:excep_XYZ})]. Here $N=12$. All the axes are dimensionless.}
    \label{fig:xyz_rt}
\end{figure*}

The similar pattern emerges for the \(iATXY\) model. Specifically, for \(\gamma =1.0\) and
with $h_a=0.5$, $h^{iATXY}_{ep}= 1.5$, and when the initial state is prepared at $h_0=2.0> h^{iATXY}_{ep}$, we observe the similar monotonic increase of \(\lambda(t)\) with \(t\) in the initial duration of the dynamics while it converges to a positive value after a certain time [see dotted lines in Fig. \ref{fig:atxy_rt} (a)]. Interestingly, we find that the convergence occurs in this case with \(N=80\) which is much lower than the \(iXY\) model and hence we carry out our investigation with \(N=100\) in Fig. \ref{fig:atxy_rt}. Unlike the \(iXY\) model, the steady-state value, \(\eta^S\), is not monotonic with $h_1$ in the quench of the $iATXY$ broken phase, irrespective of the anisotropy  and \(h_a\) as seen in  Fig. \ref{fig:atxy_rt}(b), and decreases sharply as $h_1$ approaches the exceptional point, thereby discriminating unbroken phase from the broken one through the dynamics. The investigations exhibit that the prediction of the exceptional points made by the dynamical quantities matches the exact exceptional points known for the Hamiltonian.

{\it II. Unbroken to unbroken quench.} 
Let us now move to a different picture in which the parameters of the Hamiltonian in the evolution operator are also chosen from the unbroken phase, i.e., both the initial  and quenching Hamiltonian belong to the same phase.

In the $iXY$ model, when $h_1>h^{iXY}_{ep}$, we observe that the rate function shows oscillatory behavior, with oscillations dying out as the quench parameter $h_1$ is farther away from $h_{ep}$ [see dotted lines in Fig. \ref{fig:xy_rt}(a)] while in the case of the $iATXY$ model,  $\lambda(t)$ is oscillating with very small amplitude as depicted by solid lines in Fig. \ref{fig:atxy_rt}(a). 
It is clear that for the \(iXY\) model, $\eta^S$  which does decrease initially and then increases slowly remains concave in the unbroken phase. 
The similar feature can also be observed from the behavior of $\frac{d\eta^S}{dh_1}$ which keeps on increasing in the unbroken phase, as shown in Fig. \ref{fig:xy_rt}(b). In contrast,   $\eta^S$ remains very close to $0$ in the unbroken phase for the $iATXY$ model [see  Fig. \ref{fig:atxy_rt}(b)].

\textit{Remark.}  Although the behavior from oscillatory (in the case of II) to the polynomial (in the case of I) decay is continuous, we still observe an abrupt change in long-time average of the Loschmidt rate (in the steady state domain). This corresponds to the fact that the Hamiltonian becomes nondefective in the unbroken phase from being defective in the entire broken phase after crossing the exceptional point suddenly (see Appendix \ref{appendix:defective}).

\section{Distinguishing between broken and unbroken phases from dynamics via Exact diagonalization }
\label{sec:finding_ep}

Let us now check whether the prediction made in the preceding section with the exactly solvable  models  also holds for models which cannot be solved analytically. To demonstrate that, we deal with the \(iXYZ\) models having both short- and long-range interactions given in Eqs. (\ref{eq:iXYZRT}) and (\ref{eq: iXYZRT_LR}).

To investigate the \(\mathcal{RT}\)-symmetric \(iXYZ\) models, we employ exact diagonalization 
to find the corresponding ground state, which is taken as the initial state, and find the metric. Since we require  the entire spectrum to define the metric of the Hamiltonian and we further want to study the dynamics, we cannot go beyond a certain system-sizes, which is \(N=12\). 
As discussed in the previous scenario, the initial state is prepared in the unbroken phase, i.e., $h_0 > h_{ep}$ of the corresponding model.
 
 {\it Short-range \(iXYZ\) model.} Let us first concentrate on the short-range \(iXYZ\) model. When the system is quenched to the broken phase, \(\mathcal{L}(t)\) decreases sharply at the initial period of time and then oscillates with a fixed average value during the steady state dynamics. In contrast, when the evolution operator belongs to the unbroken phase, \(\mathcal{L}(t)\) always oscillates with time having a fixed pattern and a fixed average value [see Fig. \ref{fig:xyz_rt}(a)]. The trends of \(\mathcal{L}(t)\) possibly indicates that the Loschmidt echo in the transient regime, i.e.,   \(\eta^T\) carries the signature of exceptional point or the transition between broken and unbroken phase.  As shown in Fig.~\ref{fig:xyz_rt}(b), this is indeed the case. Specifically, for a fixed values of \(\Delta\) and \(\gamma\), we observe that if \(\eta^T\)  changes the curvature from concave to convex, we can definitely conclude that the initial Hamiltonian and the final Hamiltonian are chosen from different phases, i.e.,  it is quenched from the unbroken phase to a broken phase and vice versa. In other words, at the exceptional point,  \(d\eta^T/dh_1\) surely exhibits a kink and such non-analytic behavior can also be seen if the evolution occurs  due to the change  from the unbroken phase to the broken phase [see Figs. \ref{fig:xyz_rt}(c)-\ref{fig:xyz_rt}(d)].

\begin{figure*}[!ht]
    \includegraphics[width=\linewidth]{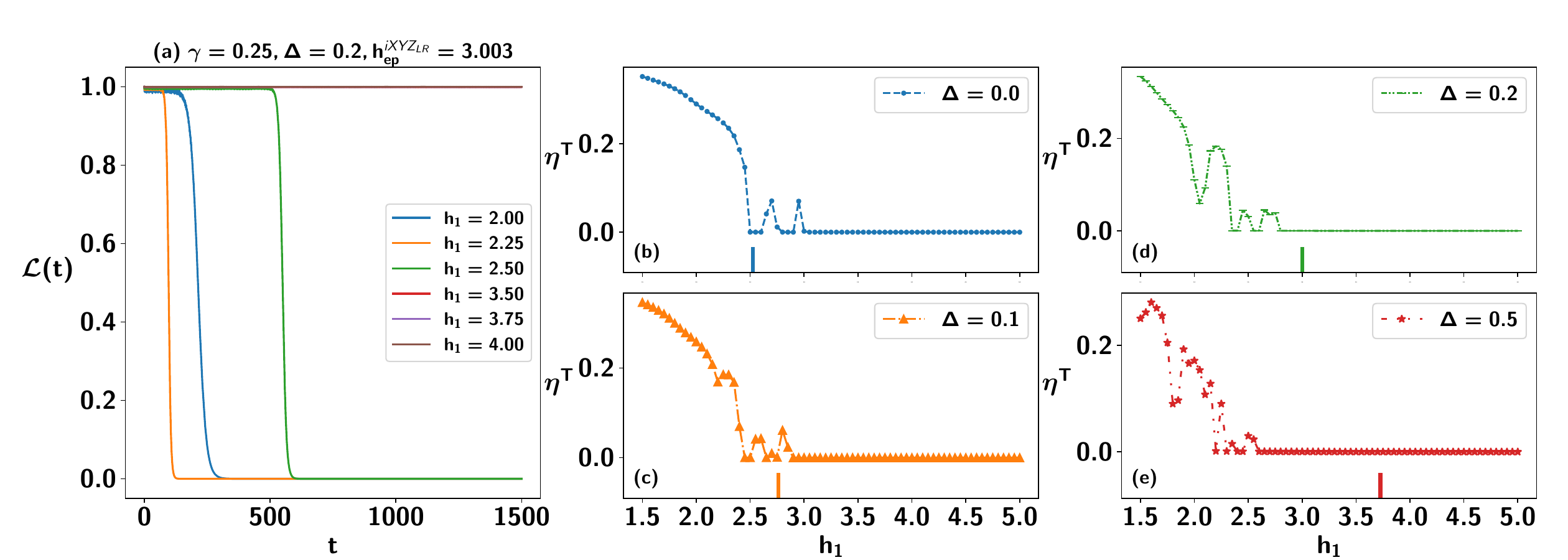}
    \caption{{\bf Predicting EP for  the long-range  $iXYZ$} with decaying parameter $\alpha=1$ and anisotropy parameter $\gamma=0.25$. The ground state is taken at $h_0=5.0 > h_{ep} = 3.003$ which is in the unbroken phase for all values of $\Delta$ considered. (a) $\mathcal{L}(t)$ (ordinate) with respect to \(t\) (abscissa). The transient time behavior is steep when the parameters of the initial and final Hamiltonians are chosen from different phases. (b)-(e) $\eta^T$ (ordinate) against the quenching magnetic field strength, $h_1$, for different $\Delta$. To calculate \(\eta^T\), $\tau_0=800$ is taken. The corresponding expected exceptional points are marked by a vertical bar [see Eq. (\ref{eq:excep_LR})]. All the axes are dimensionless.}
    \label{fig:xyz_lr_rt1}
\end{figure*}

{\it Long-range models. } The situation is much more involved when we move to the model with long-range interactions. The behavior of the Loschmidt echo clearly distinguishes two scenarios. In particular, the sharp decrease of \(\mathcal{L}(t)\) in the transient regime determines the situation when the initial state is in the unbroken phase and the Hamiltonian is tuned to the  broken phase at later time. Notice that unlike \(iXY\), \(iATXY\), and \(iXYZ\) with short-range interactions, the steep decay of \(\mathcal{L}(t)\) occurs in large \(t\).  
Such a declining trend is absent when both the initial and the final Hamiltonian are chosen from the unbroken phase. Therefore, \(\eta^T\) also carries the signature of exceptional points. As depicted in Figs. \ref{fig:xyz_lr_rt1}(b)-\ref{fig:xyz_lr_rt1}(e), we observe that when \(\eta^T\) stops fluctuating, we can conclude that the point is around the exceptional point. Specifically, we observe that for  $\Delta \le 0.2$, the short-time averaged rate function, $\eta^T$, displays a stark difference in behavior around the exceptional point although we find that with the increase of \(\Delta\), the dip observed in \(\eta^T\) is moving far from the exceptional point. 
We believe that the conclusive determination of exceptional points in the long-range case is not possible due to the finite size effect.

As shown in Figs. \ref{fig:xyz_lr_rt1}(b) and \ref{fig:xyz_lr_rt1}(c), this is indeed the case. Specifically, for fixed values of \(\Delta\) and \(\gamma\), we observe that if \(\eta^T\) changes the curvature from concave to convex, we can definitely conclude that the initial Hamiltonian and the final Hamiltonian are chosen from different phases, i.e.,  it is quenched from the unbroken phase to a broken phase and vice versa. Note that the origin of the nonanalytic behavior of the $\eta^T$ can be attributed to the change of $\mathcal{L}(t)$ from oscillatory to polynomial saturation in addition to the change from being defective to nondefective.
In other words, at the exceptional point,  \(\frac{d\eta^T}{dh_1}\) surely exhibits a kink, and such nonanalytic behavior can also be seen if the evolution occurs due to the change from the unbroken phase to the broken phase.

\section{Conclusion}
\label{sec:conclusion}

By examining the time profiles of the rate function and the long-time average of the Loschmidt echo as a function of the post quench parameters, we detect of an exceptional point (EP) in non-Hermitian systems. By using biorthogonalization method, we define a legitimate inner product in these systems. Specifically, by preparing an initial state in the unbroken phase of a non-Hermitian nearest-neighbor \(XY\) model with alternating and uniform magnetic fields, sudden quench is performed in either the broken or in the unbroken phases. We observed that the trends of rate function clearly signal the transition point in these models. More precisely, long-time behavior of the Loschmidt echo with the post quench parameter exhibits a sharp change around an exceptional point known, thereby providing a method to detect EP.  The abrupt change is attributed to the fact that Hamiltonian changes from a completely defective to a  nondefective one around the exceptional
point.

We extended this dynamical strategy for recognizing EP  in the nearest neighbor and long-range \(iXYZ\) model which can only be studied by numerical procedures.  We observed that the average Loschmidt echo in the transient domain can  signal the existence of EP, even in the presence of finite-size effects. The successful detection method based on the dynamics of quantum phase transition suggests the possibility of executing specific quantum information processing tasks during dynamics in non-Hermitian models.

\begin{acknowledgments}
We acknowledge the support from Interdisciplinary Cyber Physical Systems (ICPS) program of the Department of Science and Technology (DST), India, Grant No.: DST/ICPS/QuST/Theme- 1/2019/23. We  acknowledge the use of \href{https://github.com/titaschanda/QIClib}{QIClib} -- a modern C++ library for general purpose quantum information processing and quantum computing~\cite{qiclib} and cluster computing facility at Harish-Chandra Research Institute. This research is supported in part by the ``INFOSYS” scholarship for
senior students.

\end{acknowledgments}

\bibliography{nhquench.bib}

\appendix
\section{Diagonalization of \(iXY\) model}
\label{appendix:diag}

We now describe the diagonalization procedure for the Hamiltonian, given by
\begin{equation}
    \nonumber\bar{H}_{iXY} = \sum_{l=1}^{N}\frac{J}{4}\left [(1+i\gamma)\sigma^x_l\sigma^x_{l+1}+(1-i\gamma)\sigma^y_l\sigma^y_{l+1}\right ] - \frac{h^\prime}{2}\sigma^z_l.
    \label{eq: iXYRT_apprendix}
\end{equation}
Let us define the spin ladder operators, ($\sigma^+_l$ and $\sigma^-_l$), as
\begin{equation}
    \sigma^+_l = \frac{\sigma^x_l+i\sigma^y_l}{2}; \quad \sigma^-_l = \frac{\sigma^x_l-i\sigma^y_l}{2}\quad \forall l=1,2,\dots N.
    \label{s+-}
    \end{equation}
Thus the Hamiltonian  in terms of raising and lowering operators reduces to  
\begin{eqnarray}
    H_{iXY} = \bar{H}^{\mathcal{RT}}_{iXY} - \frac{Nh}{2} = \sum\limits_{l=1}^{N}\Bigg[\frac{i\gamma}{2}(\sigma^+_l\sigma^+_{l+1}+\sigma^-_l\sigma^-_{l+1}) \nonumber\\
    +\frac{1}{2}(\sigma^+_l\sigma^-_{l+1}+\sigma^-_l\sigma^+_{l+1}) - h\sigma^+_l\sigma^-_l\Bigg].
    \label{iXYRT1}.
\end{eqnarray}
Let us now use the highly nonlinear Jordan-Wigner transformation to represent $H_{iXY}$ in terms of fermionic creation and annihilation operators, $c_k^\dagger$ and $c_k$ respectively, with    
\begin{equation}
    c_k = e^{-i\pi\sum\limits_{l=1}^{k-1}\sigma^+_l\sigma^-_{l}}\sigma^-_k; c^\dagger_k = \sigma^+_k e^{i\pi\sum\limits_{l=1}^{k-1}\sigma^+_l\sigma^-_{l}}.
\end{equation}
In the thermodynamic limit (i.e., $N \rightarrow \infty $), the boundary term would have an infinitesimal contribution and hence ignoring the boundary term, we obtain
\begin{eqnarray}
    H_{iXY} = \sum\limits_{k=1}^{N}\Bigg[\frac{i\gamma}{2}(c^\dagger_kc^\dagger_{k+1}+c_k c_{k+1})+\nonumber \\
    \frac{1}{2}(c^\dagger_k c_{k+1}+c_k c^\dagger_{k+1}) - h c^\dagger_k c_k\Bigg].
    \label{iXYRTJW}
\end{eqnarray}
We use the Fourier transform of the fermionic operators (because of the translation invariance), given by
\begin{equation}
    a_p^\dagger = \frac{1}{\sqrt{N}}\sum\limits_{k=1}^{N}e^{-ik\phi_p} c^\dagger_k; a_p = \frac{1}{\sqrt{N}}\sum\limits_{k=1}^{N}e^{ik\phi_p} c_k, 
    \label{iXYRTFT}
\end{equation}
with $\phi_p=\dfrac{2\pi p}{N}\quad\forall p = \{-N/2,-N/2+1,\cdots -1,0,1, \cdots N/2-1, N/2\}$. Combining $\pm p$, $H_{iXY}$ is decoupled in the basis, $\{|0\rangle, a^\dagger_p a^\dagger_{-p}|0\rangle, a^\dagger_p|0\rangle,a^\dagger_{-p}|0\rangle\}$,
\begin{equation}
    \bar{H}_{iXY} = \sum\limits_{p=1}^{N/2}I\otimes I\otimes\cdots\otimes \bar{H}^p_{iXY}\otimes\dots\otimes I,
\end{equation}
with $I$ being a $4\times4$ identity matrix and
\begin{equation}
\begin{aligned}
    \bar{H}^p_{iXY}&=\left[\begin{array}{cccc}
h & -\gamma \sin \phi_p & 0 & 0 \\
\gamma \sin \phi_p  & 2\cos \phi_p-h & 0 & 0 \\
0 & 0 & \cos \phi_p & 0 \\
0 & 0 & 0 & \cos \phi_p\\
\end{array}\right], 
\end{aligned}
\end{equation}
Now, the block-diagonal form for each mode in momentum phase represented as  $\bar{H}^p_{iXY} = \hat{A}_p^\dagger H^p_{iXY} \hat{A}_p,$ where \(\hat{A}_p\) is the column vector, \((\hat{a}_{-p}^\dagger, a_{p})^T\) where $H^p_{iXY}$ is a $2\times2$ matrix represented as
\begin{equation}
    H^p_{iXY}=\left[\begin{array}{cc}
h-\cos \phi_p & -\gamma \sin \phi_p \\
\gamma \sin \phi_p & \cos \phi_p-h
\end{array}\right].
\label{Hp_ixy}
\end{equation}

\section{Defective Hamiltonian in the broken phase}

\label{appendix:defective}

In the case of the $iXY$ model, let us now show that the Hamiltonian is always defective in the broken phase. The $iXY$ Hamiltonian in the momentum space contains at least one momentum, \(p\) such that eigenvalues and the corresponding eigenvectors coalesce in the broken phase. It implies that the number of linearly independent eigenvectors is less than the dimension of the Hamiltonian which we call  \textit{defective}.  The exceptional point in the parameter corresponds to the boundary between broken and unbroken phases.

For certain values of \(h\) and \(\gamma\), \(\epsilon_p \equiv |\epsilon_p^\pm|=0\) (Eq. (\ref{eq: HiXYRT_eval})) and the corresponding eigenvectors become the same, i.e., $\frac{1}{\sqrt{2}}[1, -1]^T$ (Eq. (\ref{eq: HiXYRT_evec})). The $\epsilon_p = 0$ exists due to the following reason.  In the broken phase, $\exists$  $\phi_p\in(0,\pi)$, for which $\epsilon_p^2<0$, which is the source of complex eigenvalues.  Note that, $\lim_{\phi_p\to \pi} \epsilon_p^2 = (h+1)^2>0$ and $\lim_{\phi_p\to 0} \epsilon_p^2 = (h-1)^2>0$ for all anisotropy parameter $i\gamma$. Also, since $\epsilon_p^2$ is a continuous function of $\phi_p$, by the intermediate value theorem, there exists two momenta $\phi_{\tilde{p}_1}\in (0,\phi_p)$, and $\phi_{\tilde{p}_2}\in(\phi_p,\pi)$ such that $\epsilon_{\tilde{p}_1}^2=\epsilon_{\tilde{p}_2}^2=0$. In this case, the matrices $H^{\tilde{p}_1}_{iXY}$ and $H^{\tilde{p}_2}_{iXY}$ have only one linearly independent eigenvector i.e., $\frac{1}{\sqrt{2}}[1, -1]^T$. 
To establish the defectiveness in the spin Hamiltonian when the Hamiltonian written in momentum-space becomes defective, we consider the following scenario. Let there be four sites in the spin Hamiltonian and the corresponding momenta Hamiltonian having eigenvectors $\{|v^\pm_{p_1} \rangle, |v^\pm_{p_2} \rangle, |v^\pm_{p_3}, |v^\pm_{p_4} \rangle\}$ corresponding to $p_1$, $p_2$, $p_3$, and $p_4$ momentum blocks, respectively, i.e., the columns of the $P_p$ matrix. Hence, the eigenvectors in the spin Hamiltonian are given by $|E_i\rangle = \bigotimes|v^\pm_{p_1}\rangle |v^\pm_{p_2} \rangle |v^\pm_{p_3}\rangle |v^\pm_{p_4}\rangle $, where $i = \{1,\ldots 16\}$. It immediately implies that when two of the momentum blocks are defective,  e.g., $|v^+_{p_1}\rangle = |v^-_{p_1}\rangle$ and $|v^+_{p_3}\rangle = |v^-_{p_3}\rangle$, the corresponding spin Hamiltonian also becomes defective. This also follows as both the Jordan-Wigner and Fourier transformations are unitary transformations, they will not change the number of independent eigenvectors.

\section{Bi-orthogonalization metric}
\label{appendix:biortho}
Eigenvectors, which represent states of the  systems need to be mutually orthonormal in order to distinguish between states faithfully and maintain the probability between zero and one. In case of Hermitian systems, the observables are represented by Hermitian operators, for which, the states are orthonormal by default.

In order to check orthonormality between two vectors,  one has the inner product metric in the Hilbert space where the vectors reside. Specifically, how the dual vector is defined while finding the inner product ensures orthogonality. In the case of Hermitian quantum mechanics, the dual is the transpose of a vector after complex conjugation, and orthogonality is automatically maintained amongst eigenvectors. It is more subtle in the case of non-Hermitian systems \cite{brody_iop_2014}. 

Depending on the Hamiltonian, we must define a metric. Note first that the eigenvectors of the \(iXY\) model as shown in $P_p$  in Eq. (\ref{eq: HiXYRT_evec}) of a non-Hermitian matrix are nonorthogonal in the usual Euclidean norm. In this situation,  bi-orthogonalization \cite{Bi-ortho} can, in general, be performed for pseudo-Hermitian systems having linearly independent and complete set of eigenvectors. In particular, we define a nonsingular metric operator, $\hat{\Theta}$ which connects the left eigenvectors to the right eigenvectors for each eigenvalue, i.e., for all the vectors $|\psi\rangle$, $\langle\psi|\hat{\Theta}$ becomes its dual. This makes the eigensystem of the pseudo-Hermitian operators $H$, bi-orthogonal. The metric operator $\hat{\Theta}$ should be Hermitian and satisfy $H^\dagger\hat{\Theta} = \hat{\Theta} H$. We make the trivial choice of metric operators as
\begin{equation}
    \hat{\Theta} = (P^{-1})^\dagger P^{-1},
\end{equation}
where $P$ is the matrix with columns as the eigenvectors. This shows that $P^{-1}$ can be treated as the invertible Dyson map \cite{Dyson_map}. It allows us to treat the pseudo-Hermitian system as Hermitian in the metric space where nonorthogonal (in the Euclidean sense) basis vectors form a complete basis. Note that the matrix $P$ is invertible only in the unbroken phases, where the eigenspace is nondefective.

For our analysis, we obtain the eigenspectrum of pseudo-Hermitian $\mathcal{RT}$-symmetric Hamiltonian $H$, and obtain $\hat{\Theta}$ from the eigenvector matrix $P$ of the corresponding $H$. This is used for biorthogonalizing the Hamiltonian in the corresponding model, which makes the initial Hamiltonian Hermitian,  although during the dynamics, the quenched Hamiltonian still remains non-Hermitian.  The entire formalism enables us to study the dynamical physics of the non-Hermitian models, similarly as was done for Hermitian models in the literature \cite{Heyl_review_2018, zhou_pra_2018}. We implement such a protocol in our analytical as well as in numerical search of the exceptional point.

\end{document}